\documentclass[12pt]{iopart}

\usepackage{iopams}
\usepackage{graphicx}
\begin{document}

\title{
System size and energy dependence of $\phi$ meson production at
RHIC }

\author{J. H. Chen$^{1,2}$ for the STAR Collaboration}

\address{$^1$Nuclear Science division, Shanghai Institute of Applied
Physics, CAS, Shanghai, 201800, China} \address{$^2$Department of
Physics and Astronomy, University of California at Los Angeles,
Los Angeles, CA, 90095, USA}

\begin{abstract}
We present a system size and energy dependence of $\phi$ meson
production in Cu+Cu and Au+Au collisions at
$\sqrt{s_{NN}}=62.4$~GeV and 200~GeV measured by the STAR
experiment at RHIC. We find that the number of participant scaled
$\phi$ meson yields in heavy ion collisions over that of p+p
collisions are larger than 1 and increase with collision energy.
We compare the results with those of open-strange particles and
discuss the physics implication.

\end{abstract}
\pacs{25.75.Dw, 24.85.+p} 

\section{Introduction}
The primary aim of ultrarelativistic heavy-ion collisions is to
produce and study a state of high-density nuclear matter called
the quark-gluon plasma (QGP), the existence of which is supported
by lattice QCD calculations~\cite{LQCD-1,LQCD-2}. One of the
predicted signals for QGP formation is the enhancement of strange
particle production relative to that from light collision volume
system, p+p or p+A system. In particularly, it's argued that due
to the high production rate of $gg \rightarrow s\bar{s}$ in a QGP,
strangeness production will be increased compared to that from a
hadron gas~\cite{Strangeness-enhancement}. The subsequent
hadronization of these (anti)strange quarks results in a
significant increase in strange particle production.

On the other hand, there is an additional effect in the small
systems, e.g. p+p collisions, where a lack of available phase
space caused a suppression of strangeness
production~\cite{Canonical-suppression,Canonical-suppression-2}.
The so-called canonical suppression. However, the canonical
suppression, by definition, doesn't apply to hidden-strange
particles like $\phi$ meson. Therefore, by comparing the
normalized yields of $\phi$ mesons with those of open-strange
hadrons, one would be able to extract the information of
production dynamics.

In this paper, we will present the preliminary transverse momentum
($p_T$) distributions of $\phi$ meson in Cu+Cu collisions at
$\sqrt{s_{NN}}=62.4$~GeV and 200~GeV. The $\phi$ yield in each
$p_T$ bin was extracted from the invariant mass distributions of
$K^{+}+K^{-}$ candidates after subtraction of combinatorial
background estimated using event mixing technique. The charged
Kaons were identified through their $dE/dx$ energy loss in the
STAR Time Projection Chamber~\cite{TPC}. Details of the analysis
can be found in Ref.~\cite{STAR-phi-meson}. The $\phi$ meson
production in Au+Au collisions at the same energy will be compared
to these measurements as well.
\section{Results}
\subsection{$\phi$ meson production in Cu+Cu collisions}

\begin{figure}[htbp]
\centerline{
\includegraphics[scale=0.85]{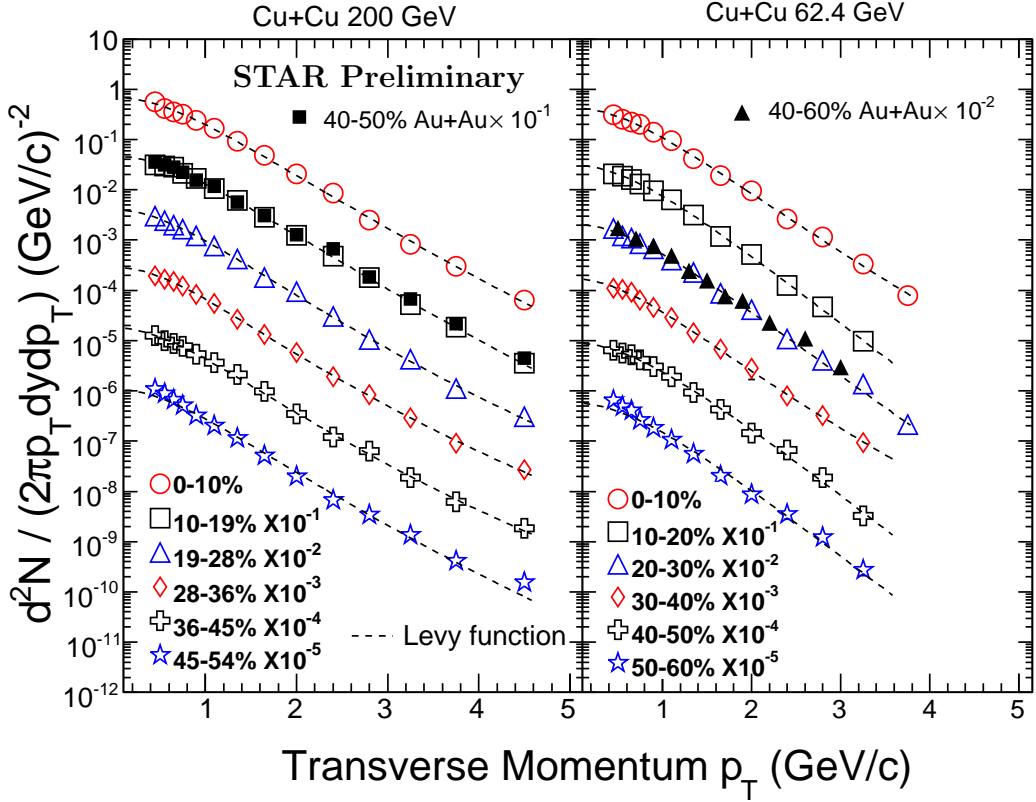}\put(-365,280){{\bf STAR
Preliminary}}} \caption{\label{fig:spectra}(color online)
Preliminary midrapidity ($|y|<0.5$) $\phi$ meson transverse
momentum spectra for various centrality classes for Cu+Cu
collisions at $\sqrt{s_{NN}}=$ 62.4 and 200~GeV. For comparison,
results of $\phi$ meson spectra from 40-50\% Au+Au collisions at
200~GeV and 40-60\% Au+Au spectra at 62.4 GeV are also shown. The
errors shown are statistical and systematic erros added in
quadrature. They are found to be within the symbol size. The
spectra are fitted to a levy function.}
\end{figure}

Figure~\ref{fig:spectra} shows the $\phi$ meson spectra in
62.4~GeV and 200~GeV Cu+Cu collisions at various centralities. The
distributions are well described by a levy function from central
to peripheral collisions. For comparison, results of $\phi$ meson
$p_T$ distributions from 40-50\% Au+Au collisions at 200 GeV and
40-60\% Au+Au collisions at 62.4 GeV are shown in the figure as
well. These centralities are chosen for similar number of
participants ($\langle N_{part} \rangle$) at corresponding
collisions energy. From the comparison, we find that, at a given
collision energy, the shape of the transverse momentum
distributions and the yield of the $\phi$ meson are scaled by the
number of participating nucleon.

\subsection{Strangeness enhancement}
\begin{figure}[htbp]
\centerline{
\includegraphics[scale=0.50]
{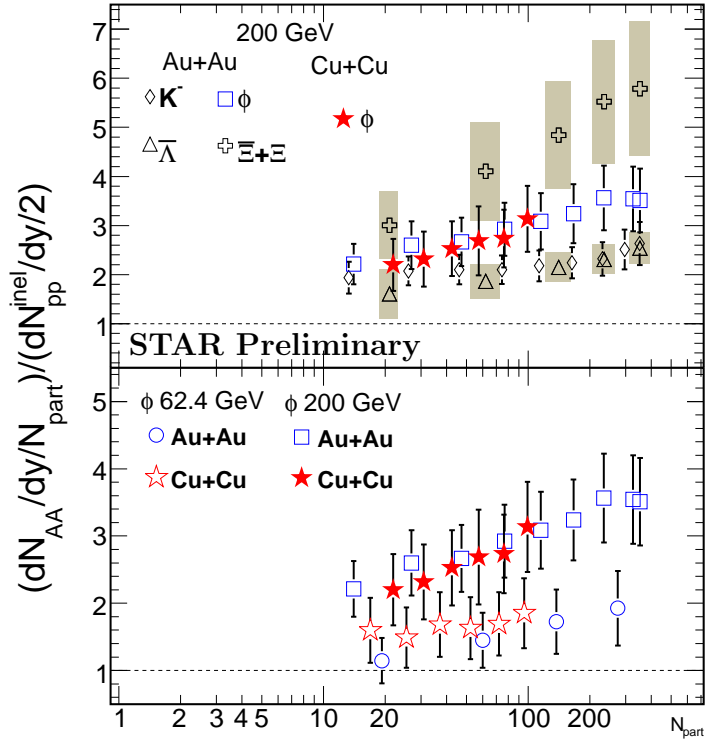} \put(-234,161){{\bf STAR
Preliminary}}} \caption{\label{fig:strange-enhance}(color online)
Upper panel: Ratio of yield of $K^-$, $\phi$, $\bar{\Lambda}$ and
$\Xi + \bar{\Xi}$ normalized to $N_{part}$ in nucleus-nucleus
collisions to corresponding yield in inelastic p+p collisions as a
function of $N_{part}$ at 200 GeV. Lower panel: Same as above for
$\phi$ meson in Cu+Cu collisions at 200 and 62.4 GeV. The p+p
collisions data from 200 GeV are from STAR~\cite{STAR-phi-meson}
and 62.4 GeV from ISR~\cite{ISR}. The error bars shown are both
statistical and systematic errors added in quadrature.}
\end{figure}

Figure~\ref{fig:strange-enhance} shows the strangeness enhancement
factor, the $N_{part}$ normalized $\phi$ yields from
nucleus-nucleus collisions relative to that from p+p collisions,
in Cu+Cu and Au+Au collisions. The results for published
open-strange hadrons~\cite{STAR-open-strange,STAR-open-strange-2}
are also shown in the figure. As one can see, the enhancement
factor for $K^{-}$, $\bar{\Lambda}$ and $\Xi + \bar{\Xi}$
increases with the number of strange valence quarks. The
enhancement in these open strange hadrons increases with collision
centrality and reaches its maximum at the most central collisions.
The enhancement of the $\phi$ meson, however, deviates from the
number of strange quark ordering. They are enhanced more than
$K^{-}$ and $\bar{\Lambda}$ but less than the $\Xi + \bar{\Xi}$.
In spite of being different particle types (meson-baryon) and
having different masses, the results for $K^{-}$ and
$\bar{\Lambda}$ are very similar in the entire centrality region
studied. This rules out a baryon-meson effect as being the reason
for the deviation of $\phi$ meson from the number of strange quark
ordering seen in Figure~\ref{fig:strange-enhance}. The observed
deviation is also not a mass effect as the enhancement in $\phi$
meson production is larger than $\bar{\Lambda}$.

The $\phi$ meson production is unlikely to be canonically
suppressed due to its $s\bar{s}$ structure. The observed
enhancement of $\phi$ meson production then is a clear indication
of dynamical effect associated with medium density being
responsible for strangeness enhancement in Au+Au collisions at 200
GeV. The observed enhancement in $\phi$ meson production being
related to medium density is further supported by the energy
dependence shown in the lower panel of
Figure~\ref{fig:strange-enhance}. The enhancement factor for
$\phi$ mesons is larger at higher collision energy, a trend
opposite to that predicted in canonical models for other strange
hadrons.

\section {Summary}

In summary, we report the preliminary results of $\phi$ meson
$p_T$ spectra in Cu+Cu collisions at $\sqrt{s_{NN}}$=62.4 and 200
GeV. Earlier published $\phi$ results from Au+Au collisions are
used for the purpose of discussions. At fixed energy, the $\phi$
mesons production seems to scaled with the $N_{part}$. The
centrality and energy dependence of the enhancement in the $\phi$
meson production clearly reflects the enhanced production of
s-quarks in a dense medium formed in high energy heavy-ion
collisions. It then indicates that the source of enhancement of
strange particle is related to the formation of a dense medium in
high energy nucleus-nucleus collisions and cannot be solely due to
canonical suppression of their production in smaller systems.

\end{document}